\documentclass[twocolumn,amsmath,amssymb]{revtex4}
\usepackage{graphicx}
\usepackage{dcolumn}
\usepackage{bm}
\begin{document}

\title{Fractional Quantum Hall Effect in Hofstadter Butterflies of Dirac Fermions}
\author{Areg Ghazaryan, Tapash Chakraborty$^\ddag$}
\affiliation{Department of Physics and Astronomy,
University of Manitoba, Winnipeg, Canada R3T 2N2}
\author{Pekka Pietil\"ainen}
\affiliation{Department of Physics/Theoretical Physics,
University of Oulu, Oulu FIN-90014, Finland}

\date{\today}
\begin{abstract}
We report on the influence of a periodic potential on the fractional quantum Hall effect (FQHE)
states in monolayer graphene. We have shown that for two values of the magnetic flux per
unit cell (one-half and one-third flux quantum) an increase of the periodic potential
strength results in a closure of the FQHE gap and appearance of gaps due to the periodic
potential. In the case of one-half flux quantum this causes a change of the ground state and 
consequently the change of the momentum of the system in the ground state. While there is also 
crossing between low-lying energy levels for one-third flux quantum the ground state does not 
change with the increase of the periodic potential strength and is always characterized by 
the same momentum. Finally, it is shown that for one-half flux quantum the emergent gaps are 
due entirely to the electron-electron interaction, whereas for the one-third flux quantum 
per unit cell these are due to both non-interacting electrons (Hofstadter butterfly pattern) 
and the electron-electron interaction.  
\end{abstract}
\maketitle

Planar, non-interacting electrons subjected to a periodic potential and a perpendicular
magnetic field was predicted to display the Hofstadter butterfly pattern in the energy spectrum
\cite{langbein}. This unique fractal pattern results from the incommensurability between two length
scales that are now present in the system: the magnetic length and the period of the external
potential. Experimental attempts to observe the pattern in semiconductor nanostructures
\cite{butterfly_old_expt} met only with limited success. While the existence of the butterfly pattern 
was indirectly confirmed in magnetotransport measurements in lateral superlattice structures,
the fractal nature of the spectrum was not observed. However, very recently, several 
experimental groups \cite{dean_13,hunt_13,geim_13} have reported observation of recursive 
patterns in Hofstadter butterfly in monolayer and bilayer graphene that was possible
solely due to the unusual electronic properties of graphene \cite{abergeletal,graphene_book}.
Although the theoretical issues of the non-interacting system in this context are largely
understood, questions remain about the precise role of electron-electron interactions in the
butterfly spectrum for graphene \cite{Apalkov_14} and even in conventional electron systems
\cite{read,daniela}. Properties of incompressible states of Dirac fermions have been established 
theoretically for monolayer graphene \cite{mono_FQHE} and bilayer graphene \cite{bi_FQHE} and the 
importance of interactions in the extreme quantum limit are well known \cite{FQHE_chapter,interaction}.
There are also experimental evidence of the FQHE states \cite{FQHE_book} in graphene 
\cite{graphene_book,FQHE_expt}. The precise role of FQHE in the fractal butterfly spectrum has remained 
unanswered however. Interestingly, in a recent experiment \cite{geim_14}, the butterfly states in the 
integer quantum Hall regime has already been explored. Understanding the effects of electron correlations 
on the Hofstadter butterfly is therefore a pressing issue. Here, we have developed the magnetic 
translation algebra \cite{mag_translation,haldane_85,read} of the FQHE states, in particular for the 
primary filling factor $\nu=\frac13$ for Hofstadter butterflies in graphene. Our results unveil a profound 
effect of the FQHE states resulting in a transition from the incompressible FQHE gap to the gap due to 
the periodic potential alone, as a function of the periodic potential strength, and also crossing of the 
ground state and low-lying excited states depending on the number of flux quanta per unit cell. 

We consider graphene in an external periodic potential
\begin{equation}
V(x,y)=V^{}_0[\mathrm{cos}(q^{}_xx)+\mathrm{cos}(q^{}_yy)],
\label{PotentialForm}
\end{equation}
where $V^{}_0$ is the amplitude of the periodic potential and $q^{}_x=q^{}_y=q^{}_0=2\pi/a^{}_0$, 
where $a^{}_0$ is the period of the external potential. Then the many-body Hamiltonian is
\begin{equation}
{\cal H}=\sum_i^{N^{}_e}\left[{\cal H}^i_B + V(x^{}_i,y^{}_i)\right] +
\tfrac12\sum_{i\neq j}^{N^{}_e}V^{}_{ij}
\label{MBHamiltonian}
\end{equation}
where ${\cal H}^i_B$ is the Hamiltonian of an electron in graphene in a perpendicular magnetic field
and the last term is the Coulomb interaction. The electron energy spectrum of graphene has twofold
valley and twofold spin degeneracy in the absence of an external magnetic field, the periodic potential
and the interaction between the electrons. We disregard the lifting of the valley degeneracy due to 
the Coulomb interaction and the periodic potential. We consider here the fully spin polarized
electron system and focus our attention on the valley $K$. The single-particle Hamiltonian ${\cal H}^{}_B$ 
is then written as
\cite{graphene_book,abergeletal,FQHE_chapter}
\begin{equation}
{\cal H}^{}_B=v^{}_F\left(\begin{array}{cc} 0 & \pi^{}_- \\ \pi^{}_+ & 0 \end{array}\right),
\label{SBHamiltonian}
\end{equation}
where $\pi^{}_\pm=\pi^{}_x\pm i\pi^{}_y$, ${\bm \pi}=\mathbf p +e\mathbf A/c$, $\mathbf p$ is the
two-dimensional electron momentum, $\mathbf A=(0,Bx,0)$ is the vector potential and $v^{}_F\approx
10^6\,\mathrm{m/s}$ is the Fermi velocity in graphene \cite{graphene_book,abergeletal}.

We consider a system of finite number $N^{}_e$ of electrons in a toroidal geometry, i.e., 
the size of the system is $L^{}_x=M^{}_xa^{}_0$ and $L^{}_y=M^{}_ya^{}_0$ ($M^{}_x$ and
$M^{}_y$ are integers) and
apply periodic boundary conditions (PBC) in order to eliminate the boundary effects. Defining the 
parameter $\alpha=\phi^{}_0/\phi$, where $\phi=Ba_0^2$ is the magnetic flux through the unit cell of 
the periodic potential and $\phi^{}_0=hc/e$ the flux quantum, we have
\begin{equation}
\frac{N^{}_s}{M^{}_xM^{}_y}=\frac1\alpha=\frac rv,
\label{FluxCond}
\end{equation}
where $N^{}_s$ is the number of magnetic flux quanta passing through the system and $r$ and $v$ are
coprime integers. The filling factor is defined as $\nu=p/q=N^{}_e/N^{}_s$, where $p$ and $q$ are 
again coprime integers. For a many-body system only the set of $\{\mathbf L^{}_{mn}/N^{}_s\}$ of the 
center-of-mass (CM) translations acts within the same Hilbert space \cite{haldane_85,FQHE_book}. Here 
$\mathbf L^{}_{mn}=mL^{}_x\hat{\mathbf x} +nL^{}_y\hat{\mathbf y}$ is a magnetic translation lattice 
vector and $(L^{}_x,L^{}_y)$ defines the magnetic translation unit cell \cite{mag_translation}.  Without 
the PBC, the Hamiltonian (\ref{MBHamiltonian}) has a symmetry of a magnetic translation of 
the CM by any periodic potential lattice vector. In order to have this symmetry in the thermodynamic 
limit, the magnetic translation of the CM by the magnetic translation lattice vector should be 
compatible with the translation by the periodic potential lattice vector \cite{read}. This compatibility
results in additional constraints on our system, that $M^{}_x$ and $M^{}_y$ are divisible by $v$.
These constraints and (\ref{FluxCond}) dictates that $N^{}_s=\kappa^{}_{x,y}M^{}_{x,y}$, where $\kappa^{}_{x,y}$ 
are integers. Therefore from the set of CM translation $\{\mathbf L^{}_{mn}/N^{}_s\}$ only those 
which are also translations by the periodic potential lattice vector will both preserve the Hilbert 
space and commute with the Hamiltonian (\ref{MBHamiltonian}).

We are seeking for the set of appropriate magnetic translations that characterizes the states of 
the Hamiltonian (\ref{MBHamiltonian}) by its momentum eigenvalues. Based on the considerations
above we search for appropriate translations as the CM translations with the translation vector
$\mathbf a^{}_p=m\beta^{}_1a^{}_0\hat{\mathbf x}+n\beta^{}_2a^{}_0\hat{\mathbf y}$,
where $\beta^{}_1$ and $\beta^{}_2$ are integers determined below. In order for these CM
translations to be diagonalized simultaneously the following condition must be satisfied
$\frac{N^{}_e\beta^{}_1\beta^{}_2}{\alpha}=\pm1,\pm2,\ldots$. By choosing 
for example $\beta^{}_2=1$ and demanding the above condition for $\beta^{}_1$, it can be shown that 
this condition is the same as the one obtained earlier by Kol and Read \cite{read}. Hence in that case 
$\beta^{}_1$ describes the degeneracy of the system. We now make the assumption that the application of 
the normal momentum operator ${\cal Q}(\mathbf Q)=\sum^{}_i\mathrm{e}^{i\mathbf Q\cdot\mathbf r^{}_i}$ to 
the many-particle state will increase its momentum by $\mathbf Q$ provided that $\mathbf Q$ is a magnetic 
translation reciprocal lattice vector. From the relation
\begin{equation}
T^\mathrm {CM}(\mathbf a^{}_p){\cal Q}(\mathbf Q^{}_{st})
=\mathrm{e}^{i\mathbf a^{}_p\cdot\mathbf Q^{}_{st}}
        {\cal Q}(\mathbf Q^{}_{st})T^\mathrm {CM}(\mathbf a^{}_p),\end{equation}
it follows that the eigenvalues of the CM translation operator will have the form $\mathrm{e}^{2\pi
i\left(\beta^{}_1ms/M^{}_x+\beta^{}_2nt/M^{}_y\right)}$, where $s$ and $t$ are integers, which 
characterize the vector $\mathbf Q^{}_{st}$ in a magnetic translation reciprocal lattice. Hence $s$ and 
$t$ are defined only modulo $M^{}_x/\beta^{}_1$ and $M^{}_y/\beta^{}_2$ respectively and there are
$M^{}_xM^{}_y/\beta^{}_1\beta^{}_2$ allowed eigenvalues. It is clear from the discussions above that $s$ 
and $t$ are related to the CM momentum of the system and also in special cases of the system size, to 
the relative momentum.

We consider the many-body states $|j^{}_1,j^{}_2,\ldots,j^{}_{N^{}_e}\rangle$ as basis states 
constructed from the single-particle eigenvectors of the Hamiltonian (\ref{SBHamiltonian})
\cite{graphene_book,abergeletal,FQHE_chapter}
\begin{equation}
\label{BaseEig}
\Psi^{}_{n,j}=C^{}_n\left(\begin{array}{c} \mathrm{sgn}(n)(-i)\varphi^{}_{|n|-1,j} \\
\varphi^{}_{|n|,j}\end{array}\right),
\end{equation}
where $C^{}_n=1$ for $n=0$ and $C^{}_n=1/\sqrt{2}$ for $n\neq0$, $\mathrm{sgn}(n)=1$ for $n>0$,
$\mathrm{sgn}(n)=0$ for $n=0$, and $\mathrm{sgn}(n)=-1$ for $n<0$. Here $\varphi^{}_{n,j}$ is the 
electron wave function in the $n$-th Landau level (LL) with the parabolic dispersion taking into account
the PBC \cite{FQHE_book,note}.
The eigenvalues of Hamiltonian (\ref{SBHamiltonian}) 
corresponding to the eigenvectors (\ref{BaseEig}) are $\epsilon^{}_n=\mathrm{sgn}(n)\hbar\omega^{}_B
\sqrt{|n|}$, where $\omega^{}_B=\sqrt{2}v^{}_F/\ell^{}_0$, $\ell^{}_0=\sqrt{c\hbar/eB}$ is the magnetic
length. The many-body state $|j^{}_1,j^{}_2,\ldots,j^{}_{N^{}_e}\rangle$ is characterized by the LL index 
$n$ and the spin of the particles. The factorization rule for CM translations
\begin{equation}T^\mathrm{CM}(\mathbf
a^{}_p)=(-1)^{N^{}_e\beta^{}_1\beta^{}_2mn/\alpha}T^\mathrm{CM}(\beta^{}_1ma^{}_0\hat{\mathbf
x})T^\mathrm{CM}(\beta^{}_2na^{}_0\hat{\mathbf y}),
\end{equation}
leads to the relations
\begin{align}
& T^\mathrm {CM}(\beta^{}_2na^{}_0\hat{\mathbf y})|j^{}_1,j^{}_2,\ldots,j^{}_{N^{}_e}\rangle
=\mathrm{e}^{i2\pi\frac{\beta^{}_2n}{M^{}_y}t}|j^{}_1,j^{}_2,\ldots,j^{}_{N^{}_e}\rangle, \\
\lefteqn{T^\mathrm {CM}(\beta^{}_1ma^{}_0\hat{\mathbf x})|j^{}_1,j^{}_2,\ldots,j^{}_{N^{}_e}\rangle}
\nonumber \\
&\quad=|j^{}_1+m\beta^{}_1\kappa^{}_x,j^{}_2+m\beta^{}_1\kappa^{}_x,\ldots,j^{}_{N^{}_e}+m\beta^{}_1
\kappa^{}_x\rangle,\end{align}
where $t=\sum^{}_ij^{}_i\, \mathrm{mod}\,(M^{}_y/\beta^{}_2)$ is the total momentum quantum number in 
the $y$ direction. Hence following the procedure outlined in Ref.~\cite{FQHE_book}, we fix the total 
momentum $t$ and construct the set ${\cal T}$ of all the $N^{}_e$ particle states with the momentum 
$t$, i.e.,
${\cal T}=\{|j^{}_1,j^{}_2,\ldots,j^{}_{N^{}_e}\rangle| \quad 0\le j^{}_i<N^{}_s, 
 \quad \sum^{}_ij^{}_i=t\ \mathrm{mod}\,\left(M^{}_y/\beta^{}_2\right)\}$.
We then divide the set ${\cal T}$ into equivalence classes by defining the states $|j'_1,j'_2,
\ldots,j'_{N^{}_e}\rangle$ and$|j^{}_1,j^{}_2,\ldots,j^{}_{N^{}_e}\rangle$ equivalent if and only if 
they are related by the rule
\begin{align}
|&j'_1,j'_2,\ldots,j'_{N^{}_e}\rangle \nonumber \\
&=|j^{}_1+ m\beta^{}_1\kappa^{}_x,j^{}_2+m\beta^{}_1\kappa^{}_x,\ldots,j^{}_{N^{}_e}+m\beta^{}_1
\kappa^{}_x\rangle.\end{align}
These equivalence classes can contain at most $M^{}_x/\beta^{}_1$ members because the momenta $j^{}_i$ 
are defined $(\mathrm{mod}\,N^{}_s)$. Let $\cal L$ be one such set represented by the state
$|j^{}_1,j^{}_2,\ldots,j^{}_{N^{}_e}\rangle$. It is clear from the construction that the members of 
this set are mapped back to the set by the translation operators $T^\mathrm {CM}(\beta^{}_1ma^{}_0
\hat{\mathbf x})$ and in fact, by any translation $T^\mathrm {CM}(\mathbf a^{}_p)$. As in the case 
of $V^{}_0=0$ \cite{FQHE_book} we can assert that the complete set of 
normalized states
\begin{align}
|(s,t)&\rangle=\frac{1}{\sqrt{|{\cal L}|}}\sum_{k=0}^{|{\cal
L}|-1}\mathrm{e}^{-i2\pi\frac{\beta^{}_1s}{M^{}_x}k} \nonumber \\
        &|j^{}_1+\beta^{}_1\kappa^{}_xk,
j^{}_2+\beta^{}_1\kappa^{}_xk,\ldots,j^{}_{N^{}_e}+\beta^{}_1\kappa^{}_xk\rangle.
\label{steigstate}
\end{align}
forms the set of the eigenstates of $T^\mathrm {CM}(\mathbf a^{}_p)$ and is used as a basis for
exact diagonalization of the Hamiltonian (\ref{MBHamiltonian}) with fixed quantum numbers $s$
and $t$. Hence the magnetic translation analysis reduces the size of the Hamiltonian matrix roughly by 
a factor of $M^{}_x/\beta^{}_1$. 

In what follows we consider the system with filling factor $\nu=1/3$. We also consider two cases 
$\alpha=1/2$ and $\alpha=1/3$. We then choose the system size based on the condition (\ref{FluxCond}) 
and the number of electrons. For $N^{}_e=4$ the system size is $M^{}_x=3$ and 
$M^{}_y=2$ for $\alpha=1/2$, and $M^{}_x=2$ and $M^{}_y=2$ for $\alpha=1/3$. For $N^{}_e=6$ 
the system size is $M^{}_x=3$ and $M^{}_y=3$ for $\alpha=1/2$, and $M^{}_x=3$ and $M^{}_y=2$ for 
$\alpha=1/3$. We evaluate the FQHE gap for two different cases when the $n=0$ LL is filled or $n=1$ LL 
is filled and we disregard the interaction between the LLs. The period of the external potential is 
taken to be $a^{}_0=20$ nm throughout.

\begin{figure}
\includegraphics[width=8cm]{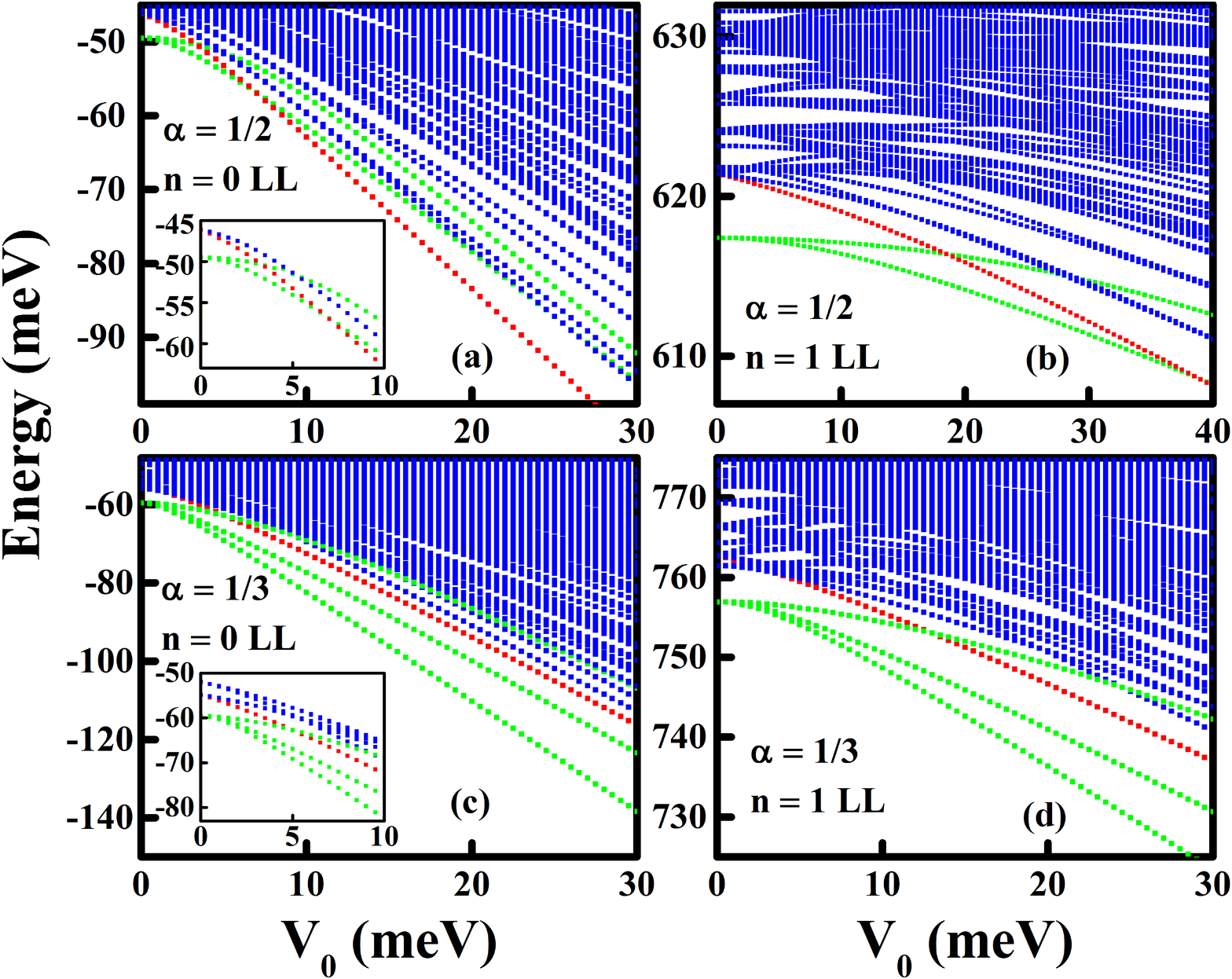}
\caption{\label{fig:Energy_Ne4} The low-lying four-electron energy levels versus
$V^{}_0$. The results are for (a) $\alpha=1/2$ and $n=0$ LL,(b)
$\alpha=1/2$ and $n=1$ LL, (c) $\alpha=1/3$ and $n=0$ LL, (d) $\alpha=1/3$ and $n=1$ LL. The triplet
ground state is shown in green and the first excited state which crosses the ground state is
shown in red. The other excited states are shown in blue. The insets show the enlarged version of the 
crossing point.}\end{figure}

In Fig. \ref{fig:Energy_Ne4} the dependence of low-lying energy levels on the amplitude of the periodic
potential $V^{}_0$ is presented for $N^{}_e=4$ and $\alpha=1/2$ and $\alpha=1/3$. Here
the levels which in the absence of the periodic potential correspond to the ground
state and become triply degenerate as $V^{}_0\to0$ are depicted in green, while the level which first
crosses those ground states is depicted in red. For $n=0$ and $n=1$ LL and for
$V^{}_0=0$, the FQHE gap is about 3.36 meV and 4 meV respectively for $\alpha=1/2$, 
and 3.69 meV and 4.513 meV respectively for $\alpha=1/3$. The difference between the gaps for
$\alpha=1/2$ and $\alpha=1/3$ comes from the fact that by fixing $\alpha$ and $a^{}_0$ we fix
the magnetic field strength $(B)$ and hence, these two cases correspond to different values of
$B$. The degeneracy of each level characterized by the CM momentum is 
$\beta^{}_1=\beta^{}_2=1$. Despite that we notice in Fig.~\ref{fig:Energy_Ne4} a,b that for 
$\alpha=1/2$ the ground state splits into two levels when the periodic potential is present. It 
should be noted that just as for $V^{}_0=0$, the spectrum as a function 
of the CM momentum has a full point symmetry of the PBC Bravais lattice. So although these three levels 
are characterized by different CM momentum, two of those are degenerate due to the PBC rectangular 
Bravais lattice. This degeneracy is not present for the cases where both $\beta^{}_1N^{}_e/M^{}_x$ 
and $\beta^{}_2N^{}_e/M^{}_y$ are integers, because as will be shown below, in these cases the relative 
momentum is a conserved quantity and the states can be characterized by both CM and relative momentum. 
When the relative momentum is a conserved quantity all three ground states correspond to both relative and 
CM momentum equal to zero and hence cannot be degenerate when the periodic potential is present. 

In the single-electron case the inclusion of a periodic potential splits the LL into $r$
subbands of equal weight \cite{Rauh}, if $\alpha=v/r$. In that case, for $\alpha=1/2$ there is
no bandgap between the two subbands. Hence the appearance of gaps for the ground and excited states for
$\alpha=1/2$ is a direct consequence of the Coulomb interaction \cite{Apalkov_14}. The 
most striking feature in Fig.~\ref{fig:Energy_Ne4} a,b for $\alpha=1/2$ is the crossing of the excited 
level with both ground states and change of the ground state at $V^{}_0\approx 7$ meV for $n=0$ LL and 
at $V^{}_0\approx 40$ meV for $n=1$ LL. The difference in the value of $V^{}_0$ where the ground state 
changes between the $n=0$ and $n=1$ LL is a direct consequence of the robustness of the FQHE state for 
$n=1$ LL compared to that of $n=0$ LL, which can be clearly seen also by the magnitude of the gaps for 
both cases above and was found earlier in graphene \cite{mono_FQHE}. These crossings and change of the 
ground state can be characterized as the the crossing of the levels with different CM momentum and also 
with different relative momentum where the relative momentum is a conserved quantity (see below). In 
Fig.~\ref{fig:Energy_Ne4} c,d and for $\alpha=1/3$ we observe similar crossing between the levels and 
closing of the FQHE gap although there is no ground state change in this case. This is related to the 
fact that we consider the system with filling factor $\nu=1/3$. For $\alpha=1/3$ this corresponds to 
the ground state of the system, which will be separated from the excited states by the inclusion of the 
periodic potential even for non-interacting electrons due to the Hofstadter gap. Even though the 
inclusion of interaction adds additional gaps to the energy spectra, as can be seen for the excited 
states in Fig.~\ref{fig:Energy_Ne4} c,d, the Hofstadter gaps are considerably larger and an increase of 
$V^{}_0$ will not result in a change of the ground state. When $\alpha=1/2$ the filling 
factor $\nu=1/3$ does not correspond to a special point because as was shown earlier \cite{Apalkov_14}, 
although for $\alpha=1/2$ there are no gaps for non-interacting electrons, interaction opens
the gaps and the highest gap is observed for $\nu=1/2$, which corresponds to the crossing points of
two subbands in the non-interacting case. Hence, this point results in the change of the ground state 
for $\alpha=1/2$, closure of the FQHE gap and, afterwards, reappearance of the gap due to the 
periodic potential.  

\begin{figure}
\includegraphics[width=8cm]{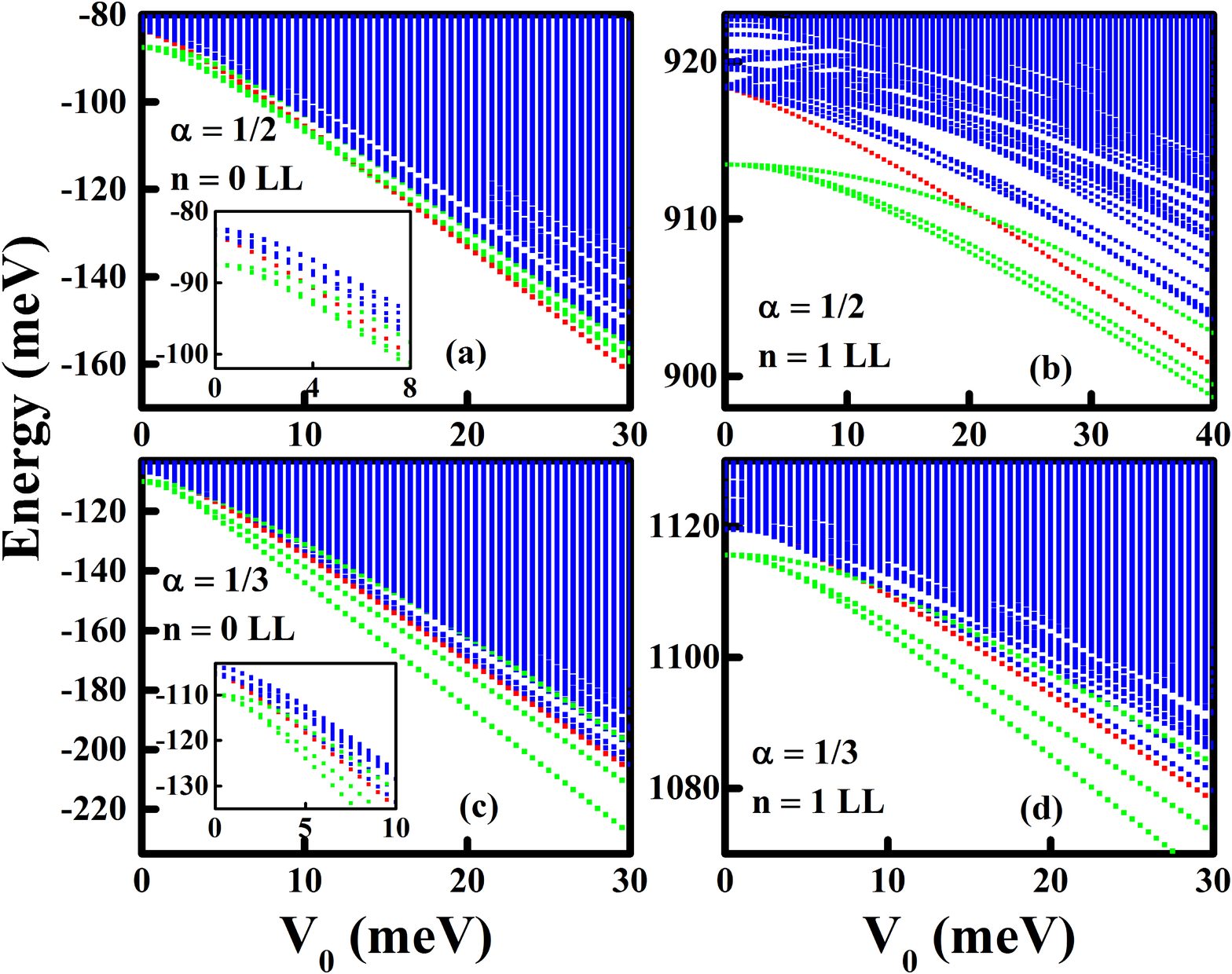}
\caption{\label{fig:Energy_Ne6} Same as in Fig.~1 but for $N^{}_e=6$.
}\end{figure}

In Fig.~2 the dependence of low-lying energy levels on $V^{}_0$ is shown for $N^{}_e=6$ and $\alpha=1/2$
and $\alpha=1/3$. The FQHE gap for $n=0$ and $n=1$ LL and $V^{}_0=0$ is $\sim$4.02 meV and $\sim$4.85 
meV respectively for $\alpha=1/2$, $\sim$4.54 meV and $\sim$5.52 meV for $\alpha=1/3$ respectively. 
Similar to the case of $N^{}_e=4$ and $\alpha=1/2$ we observe the change of the ground state
at $V^{}_0\approx16.5$ meV for $n=0$ LL and $V^{}_0\approx91$ meV for $n=1$ LL (not shown in Fig.~2 b).
For $\alpha=1/3$ we again observe a crossing between the highest ground state and the excited state, but 
do not observe any ground state change. The only difference between $N^{}_e=6$ and $N^{}_e=4$ is the 
observation of complete lifting of the degeneracy of ground state with the inclusion of the periodic 
potential, and is related to the fact that for $N^{}_e=6$ the relative momentum is a conserved quantity 
for all cases considered. It should be noted that although the value of $V_0$ at which the ground state 
changes for $\alpha=1/2$ vary considerably with the number of electrons [Fig.~1 (a,b) and Fig.~2 (a,b)], 
the value of $V^{}_0$ at which the closure of the FQHE gap appears is almost the same for both systems. 

Just as for $V^{}_0=0$, we can define the relative magnetic translations
$T^\mathrm{R}_i(N^{}_e\mathbf a^{}_p)=T^{}_i(N^{}_e\mathbf a^{}_p)T^\mathrm{CM}(\mathbf a^{}_p)$, 
which generally does not commute with the Coulomb interaction term $V^{}_{ij}$ and hence with the 
Hamiltonian (\ref{MBHamiltonian}), unless $\beta^{}_1N^{}_e/M^{}_x$ and $\beta^{}_2N^{}_e/M^{}_y$ are 
integers and, in that case, the vector $N^{}_e\mathbf a^{}_p$ is a magnetic translation lattice vector. 
These conditions are satisfied for all cases considered here, except for $N^{}_e=4$ and $\alpha=1/2$. 
When these conditions are satisfied the absolute values of the relative momentum and the CM momentum 
eigenvalues are equal and the state can be characterized both by the CM and the relative momentum 
eigenstates. As is well known \cite{haldane_85,FQHE_book}, without the periodic potential the triply
degenerate ground state is characterized by zero relative momentum. Hence for the cases when the 
conditions above are satisfied and the relative momentum is a conserved quantity, we can state
that the three gound states (depicted in green in all figures) correspond to both the relative and 
the CM momentum equal to zero for all $V^{}_0$ and the crossing observed in the figures for the
ground states result in the change of the value of relative momentum eigenstate of the ground state.

In conclusion, we have performed the magnetic translation analysis to study the effect of a
periodic potential on the FQHE in graphene for filling factor $\nu=1/3$. For $\alpha=1/2$ and 
$\alpha=1/3$, increasing the periodic potential strength $V^{}_0$ results in a closure of the FQHE gap 
and the appearance of gaps due to the periodic potential. We also find that for 
$\alpha=1/2$ this results in a change of the ground state and consequently in the change of the 
ground state momentum. For $\alpha=1/3$, despite the observation of the crossing between the low-lying 
energy levels, the ground state does not change with an increase of $V^{}_0$ and is always 
characterized by zero momentum. The difference between these two $\alpha$ s is a result of the origin 
of the gaps for the energy levels. For $\alpha=1/2$ the emergent gaps are due to the 
electron-electron interaction only, whereas for $\alpha=1/3$ these are both due to the non-interacting 
Hofstadter butterfly pattern and the electron-electron interaction.  

The work has been supported by the Canada Research Chairs Program of the Government of Canada.

\end{document}